\documentstyle[12pt,epsf]{article}

\begin{document}

\baselineskip=18.6pt plus 0.2pt minus 0.1pt

\makeatletter
\@addtoreset{equation}{section}
\renewcommand{\theequation}{\thesection.\arabic{equation}}
\renewcommand{\thefootnote}{\fnsymbol{footnote}}

\newcommand{\nn}{\nonumber}
\newcommand{\vs}[1]{\vspace*{#1}}
\newcommand{\hs}[1]{\hspace*{#1}}
\newcommand{\tr}{\mathop{\rm Tr}}
\newcommand{\p}{\partial}
\newcommand{\half}{\frac12}
\newcommand{\unit}{\hbox to 3.8pt{\hskip1.3pt \vrule height 7.4pt
    width .4pt \hskip.7pt \vrule height 7.85pt width .4pt \kern-2.4pt
    \hrulefill \kern-3pt \raise 3.7pt\hbox{\char'40}}}
\def\II{{\unit}}
\def\href#1#2{#2}
\newcommand{\vv}{\mbox{\boldmath$v$}}
\newcommand{\rr}{\mbox{\boldmath$r$}}
\newcommand{\hx}{\widehat{x}}
\newcommand{\PB}[2]{\{#1,#2\}}
\newcommand{\de}{\delta_{\epsilon}}
\newcommand{\comm}[2]{\left[#1,#2\right]}
\newcommand{\bun}[1]{\frac{1}{#1}}
\newcommand{\ve}{\varepsilon}
\newcommand{\calO}{{\cal O}}
\newcommand{\calD}{{\cal D}}
\newcommand{\calF}{{\cal F}}
\newcommand{\calG}{{\cal G}}

\makeatother

\begin{titlepage}
\title{
\hfill\parbox{4cm}
{\normalsize KUNS-1631\\{\tt hep-th/0001135}}\\
\vspace{1cm}
String Junction from Non-Commutative
\\Super Yang-Mills Theory
}
\author{
Hiroyuki {\sc Hata}\thanks{{\tt hata@gauge.scphys.kyoto-u.ac.jp}}
{} and
Sanefumi {\sc Moriyama}\thanks{{\tt moriyama@gauge.scphys.kyoto-u.ac.jp}}
\\[7pt]
{\it Department of Physics, Kyoto University, Kyoto 606-8502, Japan}
}
\date{\normalsize January, 2000}
\maketitle
\thispagestyle{empty}

\begin{abstract}
\normalsize\noindent
We construct a $1/4$ BPS soliton solution in ${\cal N}=4$
non-commutative super Yang-Mills theory to the first order
in the non-commutativity parameter $\theta_{ij}$.
We then solve the non-commutative eigenvalue equations for
the scalar fields. The Callan-Maldacena interpretation
of the eigenvalues precisely reproduces the expected string
junction picture: the string junction is tilted against the D3-branes
with angle $\theta_{ij}$.
\end{abstract}

\end{titlepage}

\section{Introduction}

Recently non-commutative Yang-Mills theory has attracted much
attention because of its origin as an effective theory of strings
\cite{CDS,DH}.
In fact, non-commutative Yang-Mills theory arises as a definite limit
of the D-brane effective theory obtained from string theory in the
presence of a constant Neveu-Schwarz 2-form $B_{\mu\nu}$ background
using point-splitting regularization.
On the other hand, the D-brane effective theory obtained from string
theory in the same situation and with Pauli-Villars regularization is
the ordinary Born-Infeld action.
Therefore, there must be a relationship between the non-commutative
Yang-Mills theory and the ordinary Born-Infeld action \cite{SW}.
Exploring this relation through the soliton solutions in both the
theories is an interesting subject.

The Born-Infeld action with a constant $B$-field background is
equivalent to the Born-Infeld action in a uniform magnetic
field, and its classical solution representing a D-string attached to
a D3-brane was analyzed in \cite{KHas}.
The result shows that the D-string tilts against the D3-brane because
of the force balance between the magnetic force and the string
tension. In the $U(2)$ non-commutative Yang-Mills case,
believing the force balance, we are led to the picture of two parallel
D3-branes with a tilted D-string suspended among them \cite{HH}.

In ref.\ \cite{HHM,Bak}, the monopole solution in non-commutative
$U(2)$ Yang-Mills theory was constructed to the first
non-trivial order in the non-commutativity parameter $\theta_{ij}$.
In order to obtain the string theory picture by the Callan-Maldacena
interpretation \cite{CM,AHas} where we identify a tube-like
configuration of a D3-brane as a D-string, we proposed in \cite{HHM}
the non-commutative eigenvalue equation for a matrix-valued fields.
{}From the eigenvalues of the Higgs scalars, we found that
the D-string tilts and the result perfectly agrees with the expected
one \cite{HH}.
Some related issues on monopoles in non-commutative Yang-Mills theory
are found in \cite{Bak} for the ADHMN construction, \cite{LYnc} for
the T-dual description, and \cite{Jia} for $U(1)$ Dirac monopoles.

\begin{figure}[htb]
\begin{center}
\leavevmode
\epsfxsize=70mm
\put(100,160){{\bf\Large $Y$}}
\put(190,92){{\bf\Large $X$}}
\epsfbox{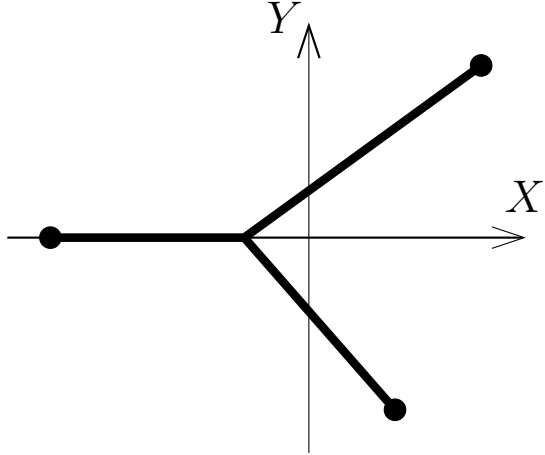}
\caption{A three-string junction. The blobs at the end of the strings
  represent D3-branes which are extended in the direction
  perpendicular to the $(X,Y)$ plane.
}
\label{XY}
\end{center}
\end{figure}

In this paper, we extend the analysis of ref.\ \cite{HHM} for the
non-commutative $U(2)$ monopole solution to the $1/4$ BPS solution in
non-commutative $U(3)$ super Yang-Mills theory.
Such solutions were constructed for the ordinary super Yang-Mills
theory in \cite{HHSsu3,HHSsuN,KO,LY}, and they gave the string
junction interpretations predicted in \cite{Sch,DM,Ber}
(see Fig.\ \ref{XY}). Here in this paper, first solving the
non-commutative BPS equations and then solving the eigenvalue
equations for the scalars, we obtain a configuration of string
junction which is tilted against the D3-branes as was expected
in \cite{HH}.

The present paper is also interesting as a testing ground of the
non-commutative eigenvalue equation, which was proposed in ref.\
\cite{HHM} and was one of the non-trivial points in the analysis
there.
In fact, the expected tilted D-string picture can never be obtained in
the $U(2)$ case if we consider only the ordinary eigenvalues of the
scalar field as a $2\times 2$ matrix.
Our result here for the $1/4$ BPS solution gives another evidence for
the validity of our non-commutative eigenvalue equation.

The rest of this paper is organized as follows.
In section 2, we shall explain the strategy for constructing
the string junction solutions in non-commutative Yang-Mills theory.
In section 3, as the first step, we solve the non-commutative monopole
equation for one of the scalars and the gauge fields
to the first order in  $\theta_{ij}$, and give a string theory
interpretation by solving the non-commutative eigenvalue equation.
Next in section 4, we solve the equation for another scalar and obtain
its eigenvalues. In section 5, we give the string junction
interpretation of the results of sections 3 and 4.
In the final section, we summarize the paper and give some
discussions.

\section{Equations for the string junction solution}

In this section we shall recapitulate some results in ordinary super
Yang-Mills theory necessary for later analysis by generalizing them to
the non-commutative case. All the results here can be found in
\cite{HHSsu3,HHSsuN,KO,LY} except that we rewrite the ordinary product
into the non-commutative star product here.

To construct the string junction solution in the four-dimensional
${\cal N}=4$ non-commutative super Yang-Mills theory with gauge group
$U(N)$, we need the bosonic part of the action consisting of the
gauge field $A_\mu$ and six scalars $X^I$ ($I=1,\ldots,6$):
\begin{eqnarray}
S=\int d^4x\tr\left(-\frac14 F_{*\mu\nu}*F_*^{\mu\nu}
-\frac12D_{*\mu}X^I*D_*^{\mu}X^I+\frac14([X^I,X^J]_*)_*^2\right) ,
\label{action}
\end{eqnarray}
with
$F_{*\mu\nu}\equiv\p_\mu A_\nu-\p_\nu A_\mu-i[A_\mu,A_\nu]_*$
and $D_{*\mu}X\equiv\p_\mu X-i[A_\mu,X]_*$.
The commutator and the square is defined by using the star product:
$[A,B]_*\equiv A*B-B*A$ and $(A)_*^2\equiv A*A$.
The star product is defined as usual by
\begin{eqnarray}
(f*g)(x)\equiv f(x)\exp\Bigl(\frac{i}{2}\theta_{ij}
\overleftarrow{\p_i}\overrightarrow{\p_j}\Bigr)g(x)
=f(x)g(x)+\frac{i}{2}\PB{f}{g}(x)+O(\theta^2) ,
\label{star}
\end{eqnarray}
where $\PB{f}{g}$ is the Poisson bracket,
\begin{eqnarray}
\PB{f}{g}(x)\equiv\theta_{ij}\,\p_if(x)\,\p_jg(x) .
\label{poisson}
\end{eqnarray}
The Gauss law constraint of this system reads
\begin{eqnarray}
D_{*i}E_{*i}=i[X^I,D_{*0}X^I]_* ,
\label{gausslaw}
\end{eqnarray}
where $E_{*i}$ is the electric field, $E_{*i}=F_{*0i}$.

The energy of this system is given by
\begin{eqnarray}
E=\int d^3x\frac12\tr\left((E_{*i})_*^2+(B_{*i})_*^2
+(D_{*0}X^I)_*^2+(D_{*i}X^I)_*^2-\frac12([X^I,X^J]_*)_*^2\right) ,
\label{energy}
\end{eqnarray}
where $B_{*i}$ is the magnetic field, $B_{*i}=\epsilon_{ijk}F_{*jk}/2$.
Hereafter we shall keep only two of the scalar fields, $X^1=X$ and
$X^2=Y$, nonvanishing. Then, using the Gauss law (\ref{gausslaw}), we
can rewrite the energy (\ref{energy}) into
\begin{eqnarray}
&&E=\int d^3x\frac12\tr\biggl\{
\Bigl(\cos\phi E_{*i}-\sin\phi B_{*i}-D_{*i}X\Bigr)_*^2
+\Bigl(\sin\phi E_{*i}+\cos\phi B_{*i}-D_{*i}Y\Bigr)_*^2\nn\\
&&\qquad\qquad\quad
+\Bigl(D_{*0}X+i\sin\phi\left[X,Y\right]_*\Bigr)_*^2
+\Bigl(D_{*0}Y-i\cos\phi\left[X,Y\right]_*\Bigr)_*^2\biggr\}\nn\\
&&\qquad\qquad\quad
+(Q_X+M_Y)\cos\phi+(Q_Y-M_X)\sin\phi\nn\\
&&\phantom{E}
\ge (Q_X+M_Y)\cos\phi+(Q_Y-M_X)\sin\phi ,
\label{bound}
\end{eqnarray}
where $\phi$ is an arbitrary parameter, and we have
$Q_X=\int dS_i\tr(E_{*i}*X)$ and $M_X=\int dS_i\tr(B_{*i}*X)$,
and similarly for $Q_Y$ and $M_Y$.
{}From (\ref{bound}) we obtain the classical equations as the condition
for saturating the lower bound. We can put $\phi=0$ without loss of
generality since $\phi$ can be varied by a rotation in the $(X,Y)$
plane. Therefore the equations to be solved are
\begin{eqnarray}
&&D_{*i}X=E_{*i} ,
\label{DiX}\\
&&D_{*i}Y=B_{*i} ,
\label{mpeq}\\
&&D_{*0}X=0 ,
\label{D0X}\\
&&D_{*0}Y=i\left[X,Y\right]_* ,
\label{D0Y}\\
&&D_{*i}D_{*i}X=\left[Y,\left[Y,X\right]_*\right]_* ,
\label{jcteq}
\end{eqnarray}
where the last equation (\ref{jcteq}) is the Gauss law
(\ref{gausslaw}) with (\ref{D0X}) and (\ref{D0Y}) substituted.

Since we are interested in static solutions, we will drop the
time-dependence of all the fields. Then, eqs.\ (\ref{DiX}),
(\ref{D0X}) and (\ref{D0Y}) are automatically satisfied by putting
$A_0=-X$.
The remaining equations we have to solve are eqs.\ (\ref{mpeq}) and
(\ref{jcteq}), which we call non-commutative monopole equation and
non-commutative Gauss law, respectively.

In the commutative limit $\theta=0$, eq.\ (\ref{mpeq}) reduces to the
ordinary BPS monopole equation which was solved in the seminal papers
\cite{Bog,PS} by adopting the spherical symmetry ansatz:
\begin{eqnarray}
A_i^0=-\epsilon_{ijk}\hx_jT_k(K(\xi)-1)/r ,\qquad
Y^0=-\hx_iT_iH(\xi)/r ,
\label{mpsol0}
\end{eqnarray}
where the superscripts $0$ on $A_i$ and $Y$ denote that they are the
$0$-th order solution in $\theta$.
The dimensionless quantities $\hx_i$ and $\xi$ are defined by
$\hx_i\equiv x_i/r$ and $\xi\equiv Cr$ using an arbitrary constant $C$
with mass dimension.
The matrices $T_i$ ($i=1,2,3$) are an embedding of $SU(2)$ into the
$U(N)$ group: $\left[T_i,T_j\right]=i\epsilon_{ijk}T_k$.
In the case of the maximal embedding to $U(3)$, the explicit forms of
$T_i$ are
\begin{equation}
T_1=\bun{\sqrt{2}}\pmatrix{0&1&0\cr 1&0&1\cr 0&1&0} ,\quad
T_2=\frac{i}{\sqrt{2}}\pmatrix{0&-1&0\cr 1&0&-1\cr 0&1&0} ,\quad
T_3=\pmatrix{1&0&0\cr 0&0&0\cr 0&0&-1} .
\label{Ti}
\end{equation}
Putting this ansatz into (\ref{mpeq}) with $\theta=0$,
we obtain the equations for $K$ and $H$,
\begin{eqnarray}
\calD K=-HK ,\qquad\calD H=H+1-K^2 ,
\label{HK}
\end{eqnarray}
where $\calD$ denotes the Euler derivative with respect to $\xi$,
$\calD\equiv \xi\left(d/d\xi\right)$.
Eq.\ (\ref{HK}) are solved to give
\begin{eqnarray}
K=\xi/\sinh\xi ,\qquad H=\xi/\tanh\xi-1 .
\end{eqnarray}
The behaviors of $K$ and $H$ in the asymptotic region $\xi\to\infty$
are
\begin{equation}
K=O\!\left(e^{-\xi}\right) ,\qquad
H=\xi -1 + O\!\left(e^{-\xi}\right) .
\label{KHasympt}
\end{equation}

As for the Gauss law (\ref{jcteq}) with $\theta=0$, the
spherical symmetry ansatz,\footnote{
In the case of the maximal embedding to $SU(3)$, eq.\  (\ref{jctsol0})
is the most general spherically symmetric form for $X$ since $T_i$ and
$T_{ij}$ span the whole $SU(3)$. However, for $SU(N)$ with $N\ge 4$,
there are other spherically symmetric terms using
the symmetric traceless products of $T_i$'s.
}
\begin{eqnarray}
X^0=\bun{r}\left(
\hx_iT_i P(\xi)+\hx_i\hx_jT_{ij}\frac{Q(\xi)}{\xi} \right) ,
\label{jctsol0}
\end{eqnarray}
was considered in \cite{HHSsu3} where
$T_{ij}\equiv\{T_i,T_j\}-\delta_{ij}T_0/3$ with
$T_0\equiv\{T_i,T_i\}$. In the particular case of the maximal
embedding into $U(N)$, we have $T_0=(N^2-1)\unit/2$.
The Gauss law with $\theta=0$ under the ansatz (\ref{jctsol0})
was solved to give
\begin{eqnarray}
&&P = -\alpha H ,\\
&&Q = -\beta\left(2H^2+H-1+K^2\right) ,
\label{N}
\end{eqnarray}
where $\alpha$ and $\beta$ are arbitrary constants.\footnote{
Our $(\alpha,\beta)$ is related to that in ref.\ \cite{HHSsu3} by
$(\alpha,\beta)_{\mbox{\scriptsize ref.\cite{HHSsu3}}}=(-4\alpha,(8/3)\beta)$.
}

In the next two sections we shall solve the non-commutative monopole
equation (\ref{mpeq}) and the non-commutative Gauss law
(\ref{jcteq}) by the $\theta$ expansion. First we shall expand them to
the first non-trivial order in $\theta$ and solve them by adopting
(\ref{mpsol0}) and (\ref{jctsol0}) as the zero-th order solution.

\section{Non-commutative $U(3)$ monopole}
In this section, we shall solve the non-commutative monopole equation
(\ref{mpeq}) to the first order in $\theta$ and evaluate
the eigenvalue of the scalar $Y$ for the brane interpretation.
By expanding (\ref{mpeq}) to the first order in $\theta$, we get
\begin{eqnarray}
\half\epsilon_{ijk}\Bigl(\p_jA_k^1-\p_kA_j^1
-i[A_j^0,A_k^1]-i[A_j^1,A_k^0]\Bigr)
-\Bigl(\p_iY^1-i[A_i^0,Y^1]-i[A_i^1,Y^0]\Bigr)\nn\\
=-\half\epsilon_{ijk}\{A_j^0,A_k^0\}
+\half\{A_i^0,Y^0\}-\half\{Y^0,A_i^0\} ,
\label{mpeq1}
\end{eqnarray}
where $A_i^1$ is the $O(\theta^1)$ part of $A_i$, namely,
$A_i=A_i^0 + A_i^1 +\ldots$, and similarly for $Y$.
Using the zero-th order solution (\ref{mpsol0}), we find that the
right-hand-side (RHS) of (\ref{mpeq1}) is given as a sum of six terms
with the following tensor structures concerning $\theta$, the open
index $i$ and the $U(N)$ Lie algebra matrix:
\begin{equation}
\theta_iT_0 ,\quad
\theta_j\hx_i\hx_jT_0 ,\quad
\theta_jT_{ij} ,\quad
\theta_i\hx_j\hx_kT_{jk} ,\quad
\theta_j\hx_i\hx_kT_{jk} ,\quad
\theta_j\hx_j\hx_kT_{ik} ,
\label{mprhs}
\end{equation}
where we have used $\theta_i\equiv\epsilon_{ijk}\theta_{jk}/2$.
The coefficient of each quantity in (\ref{mprhs}) is a polynomial of
$H$ and $K$ divided by $r^4$.
Apparently, there is another tensor structure
$\epsilon_{ijk}\epsilon_{lmn}\theta_l\hx_j\hx_mT_{kn}$
which can appear on the RHS of (\ref{mpeq1}).
However, it is not independent due to the identities,
\begin{eqnarray}
&&\epsilon_{ljk}\hx_i\hx_l
+\epsilon_{ilk}\hx_j\hx_l+\epsilon_{ijl}\hx_k\hx_l=\epsilon_{ijk} ,
\nn\\
&&\epsilon_{ljk}T_{il}+\epsilon_{ilk}T_{jl}+\epsilon_{ijl}T_{kl}=0 .
\label{identity}
\end{eqnarray}
Using either of them, we can show that
$\epsilon_{ijk}\epsilon_{lmn}\theta_l\hx_j\hx_mT_{kn}
=-\theta_jT_{ij}
-\theta_i\hx_j\hx_kT_{jk}
+\theta_j\hx_i\hx_kT_{jk}
+\theta_j\hx_j\hx_kT_{ik}$.

To solve (\ref{mpeq1}), let us adopt the generalized spherical
symmetry  ansatz \cite{HHM} for $A_i^1$ and $Y^1$. Here the
generalized spherical symmetry implies the covariance under the
combined rotations of $\theta_i$ as well as of $x_i$ and $T_i$.
Noting that all the terms of (\ref{mprhs}) are given using
either $T_0$ or $T_{ij}$ for the matrix structure and even numbers of
$\hx_i$, we see that the ansatz for $A_i^1$ and $Y^1$ should
be given by using $T_0$ or $T_{ij}$ and odd numbers of $\hx_i$.
For the gauge field $A_i^1$, at first sight the following
seven tensor structures are possible:
$\epsilon_{ijk}\theta_j\hx_kT_0$,
$\epsilon_{ijk}\theta_j\hx_lT_{kl}$,
$\epsilon_{ijk}\theta_l\hx_jT_{kl}$,
$\epsilon_{jkl}\theta_j\hx_kT_{il}$,
$\epsilon_{ijk}\theta_j\hx_k\hx_l\hx_mT_{lm}$,
$\epsilon_{ijk}\theta_l\hx_j\hx_l\hx_mT_{km}$
and $\epsilon_{jkl}\theta_j\hx_i\hx_k\hx_mT_{lm}$.
However, due to the identities (\ref{identity}),
there are two linear relations among them.
Therefore, taking all the independent tensor
structures into account, the ansatz is given as follows:
\begin{eqnarray}
&&A_i^1=\bun{r^3}\Bigl(
\epsilon_{ijk}\theta_j\hx_kT_0A(\xi)
+\epsilon_{ijk}\theta_j\hx_lT_{kl}B(\xi)
+\epsilon_{jkl}\theta_j\hx_kT_{il}C(\xi)
\nn\\
&&\qquad\qquad
+\epsilon_{ijk}\theta_j\hx_k\hx_l\hx_mT_{lm}D(\xi)
+\epsilon_{ijk}\theta_l\hx_j\hx_l\hx_mT_{km}E(\xi)\Bigr) ,
\label{Ai1}\\
&&Y^1=\bun{r^3}\Bigl(
\theta_i\hx_iT_0U(\xi)
+\theta_i\hx_jT_{ij}V(\xi)
+\theta_i\hx_i\hx_j\hx_kT_{jk}W(\xi)\Bigr) .
\label{Y1}
\end{eqnarray}
Putting this ansatz into the LHS of (\ref{mpeq1}), we obtain the
following system of linear differential equations with inhomogeneous
terms:
\begin{eqnarray}
&&\calD A-2A-U=\Bigl(-H^2K+H(K-1)^2-K(K-1)^2\Bigr)/6 ,
\label{tT0}\\
&&\calD(-A-U)+4A+4U=\Bigl(H^2K-H(K-1)(K-3)+(K-1)^3\Bigr)/6 ,
\label{txxT0}\\
&&-\calD C+2C+K(-B+C-V)+H(-2C)=(HK+K-1)(H+K-1)/2 ,
\label{tTij}\\
&&\calD(B-C+D)-2B+2C-2D-V-W+K(-2B+2C+V)+H(-C)\nn\\
&&\hs{4cm}=\Bigl(H^2K+H(K-1)-(K-1)^2(K+1)\Bigr)/2 ,\\
&&\calD(C-V)+B-2C+E+3V+K(-B-2C-2D+V)+H(B+C)\nn\\
&&\hs{4cm}=\Bigl(-H^2K-H(K-1)+(K-1)^2(K+1)\Bigr)/2 ,\\
&&\calD(-B+C-E)+3B-3C+2E+V+K(B-C-V-2W)+H(-B+2C-E)\nn\\
&&\hs{4cm}=-(HK+2K-2)(H+K-1)/2 ,\\
&&\calD(-D+E-W)+4D-3E+4W+K(2D-3E+2W)+HE=0 .
\label{txxTij}
\end{eqnarray}
They are respectively the coefficients of the six structures of
(\ref{mprhs}) and of $\theta_j\hx_i\hx_j\hx_k\hx_l T_{kl}$ (the last
one is missing on the RHS of (\ref{mpeq1})).
The first two differential equations (\ref{tT0}) and (\ref{txxT0})
for $A$ and $U$ are the $U(1)$ parts of the monopole equation
(\ref{mpeq1}) and decouple from the rest. These are exactly what we
solved in the $U(2)$ case \cite{HHM}:
\begin{eqnarray}
&&A = \bun{12}(K-1)(2H-K+1) ,\\
&&U = 0 .
\end{eqnarray}
The rest of the equations (\ref{tTij})--(\ref{txxTij}) is very
complicated and seems hard to solve at first sight.
However, we can solve them by assuming that the solutions are given as
polynomials of $H$ and $K$.
This polynomial assumption is possible due to the property
(\ref{HK}) implying that a polynomial of $H$ and $K$ acted by $\calD$
is again a polynomial of them. Concretely, we assume that
\begin{eqnarray}
\calO=\sum_{n=0}^{N_{\rm max}}\sum_{m=0}^{M_{\rm max}}
\calO_{nm}H^nK^m ,
\label{polynom}
\end{eqnarray}
for the unknown functions $\calO=B,C,\ldots$ with suitably large
$N_{\rm max}$ and $M_{\rm max}$.
Then, using the property (\ref{HK}), the differential equations
(\ref{tTij})--(\ref{txxTij}) are reduced into a set of linear
algebraic equations for the coefficients $\calO_{nm}$.
This set of algebraic equations is easily solved to give
\begin{eqnarray}
&&B= -\bun{4}+\bun{4}HK +\bun{4}K^2
+z \calF ,\label{B}\\
&&C= \bun{4} -\frac{3}{4}K -\bun{2}HK+\frac{3}{4}K^2
+ \bun{2}HK^2 -\bun{4}K^3 ,
\\
&&D= \frac{7}{8}-\bun{8}H - \bun{4}K-\frac{3}{4}HK
-\frac{7}{8}K^2 +\bun{4}HK^2 + \bun{4}K^3 + z \calG ,\\
&&E= 0 ,\\
&&V= \bun{2} + \bun{4}H -\bun{4}K-\frac{3}{4}HK
- \bun{2}K^2 + \bun{4}HK^2 + \bun{4}K^3 - z \calF ,\\
&&W= -\frac{7}{8} + \bun{8}H + \bun{4}K + \frac{3}{4}HK
+ \frac{7}{8}K^2 - \bun{4}HK^2 - \bun{4}K^3 - z \calG ,
\label{W}
\end{eqnarray}
with
\begin{eqnarray}
&&\calF=(-1+H+2H^2)K+K^3 ,\nn\\
&&\calG=\frac32(1+H)+(1-H-2H^2)K-\frac32(1+2H)K^2-K^3 .
\end{eqnarray}
The solution (\ref{B})--(\ref{W}) does not contain any singularity at
the origin $r=0$ which invalidates the integration by parts necessary
for rewriting the energy (\ref{energy}) into (\ref{bound}).

In (\ref{B})--(\ref{W}), the $\calF$ and $\calG$ terms multiplied by
an arbitrary parameter $z$ are a homogeneous solution, namely a
solution to the zero-mode equation in the ordinary monopole equation.
As we shall see later, this homogeneous solution corresponds to the
separation of the monopole centers in the $\theta_i$
direction.
Besides this zero-mode there is no spherically symmetric homogeneous
solution of our interest related to the moduli of the ordinary
monopole solution.

Here we should mention the gauge freedom of our solutions.
Apparently the set of differential equations
(\ref{tT0})--(\ref{txxTij}) is underdeterminant since the number of
unknown functions is one more larger than that of the equations.
However, this problem is resolved by noticing that there is a freedom
of local gauge transformation,
$\delta_\ve A_i=D_{*i}\ve$ and
$\delta_\ve X^I=-i[X^I,\ve]_{*}$,
with $\ve=\epsilon_{ijk}\theta_i\hx_j\hx_lT_{kl}\,L(\xi)/r^2$,
which keeps the generalized spherical symmetry.
In the solution (\ref{B})--(\ref{W}) we have put $E=0$ by using the
freedom of $L(\xi)$.

\begin{figure}[hbt]
\begin{center}
\leavevmode
\epsfxsize=12cm
\epsfbox{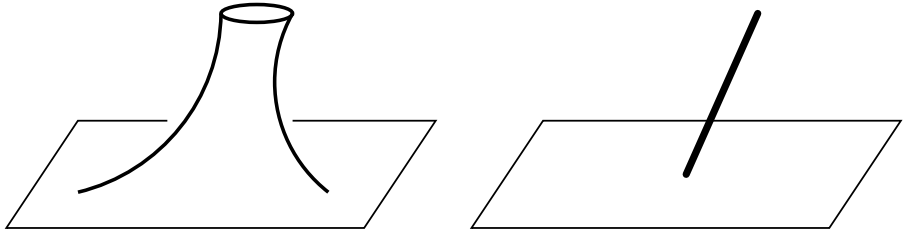}
\caption{The Callan-Maldacena interpretation. A tube-like
  deformation of a D3-brane represented by the eigenvalue of the
  scalar (left) is regarded as a D-string attached to the D3-brane
  (right).
}
\label{figCM}
\end{center}
\end{figure}

Having obtained the classical solution to $O(\theta)$, let us next
solve the non-commutative eigenvalue equation for the scalar
$Y$ in the $U(3)$ case with the maximal $SU(2)$ embedding, and then
give the brane interpretation following Callan and Maldacena \cite{CM}
(see Fig.\ \ref{figCM}).
The non-commutative eigenvalue equation for a matrix-valued function
$M$ proposed in \cite{HHM} is as follows:
\begin{eqnarray}
M*\vv=\lambda*\vv ,
\label{Mv=lv}
\end{eqnarray}
where $\lambda$ is the eigenvalue and $\vv$ is the eigenvector.
Expanding the matrix, the eigenvalue and the eigenvector as
\begin{eqnarray}
M=M^0+M^1,\quad\lambda=\lambda^0+\lambda^1,\quad\vv=\vv^0+\vv^1 ,
\label{0+1}
\end{eqnarray}
and plugging them into the eigenvalue equation (\ref{Mv=lv}),
the $O(\theta^1)$ part $\lambda^1$ of the eigenvalue is given as
\cite{HHM},
\begin{eqnarray}
\lambda^1=\frac{i}{2}\vv^{0\dagger}\PB{M^0-\lambda^0\unit}{\vv^0}
+\vv^{0\dagger}M^1\vv^0 ,
\label{eigenvalue}
\end{eqnarray}
where $\vv^0$ is normalized, $\vv^{0\dagger}\vv^0=1$.

The three zero-th order eigenvalues of the scalar $Y$ are
\begin{eqnarray}
\lambda_Y^0=-\frac{H}{r}\pmatrix{1\cr 0\cr -1}
\sim\left(-C+\bun{r}\right)\pmatrix{1\cr 0\cr -1} ,
\label{lY0}
\end{eqnarray}
where the last expression is the asymptotic ($r\to\infty$) form
obtained by dropping the exponentially decaying terms $O(e^{-\xi})$.
Applying (\ref{eigenvalue}) to the scalar $Y$ with $Y^1$ given by
(\ref{Y1}), the $O(\theta^1)$ part of the eigenvalues are
\begin{eqnarray}
&&\lambda_Y^1=\frac{\theta\cdot\hx}{r^3}\pmatrix{
H/2 + 4U + (2/3)(V+W) \cr
H + 4U -(4/3)(V+W)    \cr
H/2 + 4U + (2/3)(V+W)
}
\sim\frac{\theta\cdot\hx}{4 r^3}\pmatrix{
(3 - 4z)\xi - 4 \cr
(2 + 8z)\xi     \cr
(3 - 4z)\xi - 4
} .
\label{lY1}
\end{eqnarray}
As in the $U(2)$ case of ref.\ \cite{HHM}, the eigenvalue
$\lambda_Y^1$ is singular at the origin $r=0$ though the classical
solution is regular there. Therefore, we shall restrict our brane
analysis to the asymptotic region $r\to\infty$.
Summing (\ref{lY0}) and (\ref{lY1}), we obtain the total eigenvalues
$\lambda_Y=\left(\lambda_Y^{(+)},\lambda_Y^{(0)},\lambda_Y^{(-)}
\right)^T$:
\begin{eqnarray}
&&\lambda_Y^{(\pm)}
\sim\mp C\pm\frac{1}{r}
+\frac{\theta\cdot\hx}{4r^3}\Bigl((3-4z)\xi-4\Bigr)
\nn\\
&&\phantom{\lambda_Y^{(\pm)}}
=\mp C \pm\left|x_i
+\theta_i\left(\pm\left(\frac14+z\right)C
+\lambda_Y^{(\pm)}\right)\right|^{-1} ,
\label{lY(pm)}\\[5pt]
&&\lambda_Y^{(0)} \sim
\frac{\theta\cdot\hx}{2r^3}(1+4z)\xi
=\sum_{\pm}\mp\left|x_i
\pm\theta_i\left(\frac14+z\right)C\right|^{-1} .
\label{lY(0)}
\end{eqnarray}

\begin{figure}[hbt]
\begin{center}
\leavevmode
\epsfxsize=80mm
\put(-17,25){D3($\lambda_Y^{(+)}$)}
\put(-17,71){D3($\lambda_Y^{(0)}$)}
\put(-17,130){D3($\lambda_Y^{(-)}$)}
\put(205,25){$-C$}
\put(205,140){$C$}
\put(135,55){D-string}
\put(100,170){{\Large $Y$}}
\put(220,92){{\Large $x_i$}}
\put(135,15){{$(3/4-z)C\theta_i$}}
\put(31,92){{$(-1/4-z)C\theta_i$}}
\put(122,73){{$(1/4+z)C\theta_i$}}
\put(20,150){{$(-3/4+z)C\theta_i$}}
\epsfbox{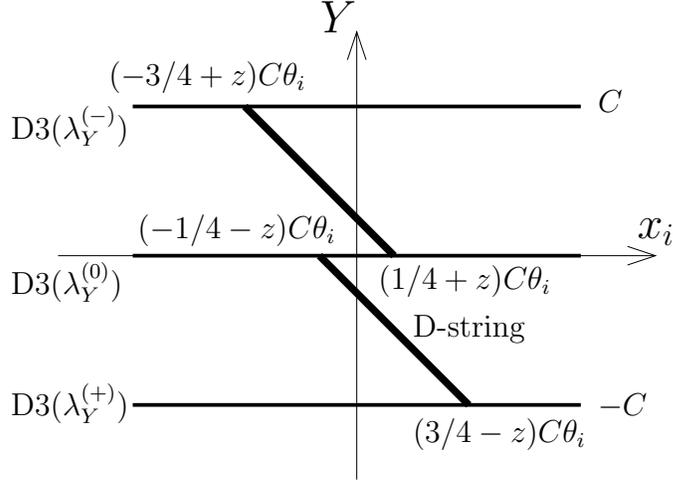}
\caption{The D-string picture obtained from the asymptotic behavior
  of the eigenvalues $\lambda_Y$. The endpoint $x_i$ coordinates of
  the D-strings are given on each D3-brane.
}
\label{Y}
\end{center}
\end{figure}

{}From (\ref{lY(pm)}) and (\ref{lY(0)}) we can read off the brane
configuration representing tilted D-strings suspended among
three parallel D3-branes as depicted in Fig.\ \ref{Y}.
First, eq.\ (\ref{lY(pm)}) implies that, for a given value of
$\lambda_Y^{(\pm)}$, the corresponding worldvolume coordinate $x_i$ is
located on a sphere with its center at
$x_i^{{\rm C}(\pm)}=-\theta_i\left(\pm\left(1/4+z\right)C
+\lambda_Y^{(\pm)}\right)$. The tilt angle of the D-strings is read
off as $-\theta_i$, and the $x_i$ coordinates of the points where the
D-strings stick to D3-branes are given as $x_i^{{\rm C}(\pm)}$
corresponding to $\lambda_Y^{(\pm)}=\mp C$.

Next, from $\lambda_Y^{(0)}$ of (\ref{lY(0)}) we see that the middle
D3-brane, which was totally flat in the commutative case $\theta=0$
(recall eq.\ (\ref{lY0})), now suffers a deformation with centers at
$x_i=\mp(1/4+z)C\theta_i$, which
we interpret as the coordinates where the D-strings meet the
middle D3-brane. This interpretation is consistent with the D-string
picture obtained above from $\lambda_Y^{(\pm)}$.
Since, when $\theta=0$, the middle D3-brane was completely flat and
did not have any parts identifiable as D-strings, it is impossible to
read off the tilt angle of the D-strings from $\lambda_Y^{(0)}$
in the present $O(\theta)$ analysis.

As seen from Fig.\ \ref{Y}, the parameter $z$ corresponds to the
relative separation of the two D-strings (namely, the separation of
the two monopole centers).
Note that the two D-strings are smoothly connected to each other only
for a special value of $z$, $z=-1/4$.

\section{Non-commutative Gauss law}

Having solved the non-commutative monopole equation (\ref{mpeq}),
let us turn to the non-commutative Gauss law (\ref{jcteq}).
We shall consider the case of $U(3)$ with the maximal embedding of
$SU(2)$.
Since the procedure for solving the Gauss law is quite
similar to that for the monopole equation (\ref{mpeq}), we shall
be brief.
Expanding (\ref{jcteq}) to the first order in $\theta$, we have
\begin{eqnarray}
&&\p_i\Bigl(\p_iX^1-i[A_i^0,X^1]\Bigr)
-i\Bigl[A_i^0,\p_iX^1-i[A_i^0,X^1]\Bigr]
-\Bigl[Y^0,[Y^0,X^1]\Bigr]\nn\\
&&\quad=\phantom{}-\p_i\biggl(-i[A_i^1,X^0]
+\half\Bigl(\{A_i^0,X^0\}-\{X^0,A_i^0\}\Bigr)\biggr)\nn\\
&&\quad\phantom{=}+i\Bigl[A_i^1,\p_iX^0-i[A_i^0,X^0]\Bigr]
+i\biggl[A_i^0,-i[A_i^1,X^0]
+\half\Bigl(\{A_i^0,X^0\}-\{X^0,A_i^0\}\Bigr)\biggr]\nn\\
&&\quad\phantom{=}-\half\biggl(\Bigl\{A_i^0,\p_iX^0-i[A_i^0,X^0]\Bigr\}
-\Bigl\{\p_iX^0-i[A_i^0,X^0],A_i^0\Bigr\}\biggr)\nn\\
&&\quad\phantom{=}+i\Bigl[Y^1,-i[Y^0,X^0]\Bigr]+i\biggl[Y^0,-i[Y^1,X^0]
+\half\Bigl(\{Y^0,X^0\}-\{X^0,Y^0\}\Bigr)\biggr]\nn\\
&&\quad\phantom{=}-\half\biggl(\Bigl\{Y^0,[Y^0,X^0]\Bigr\}
-\Bigl\{[Y^0,X^0],Y^0\Bigr\}\biggr) .
\label{jcteq1}
\end{eqnarray}

Due to the monopole equation (\ref{mpeq}) and the Bianchi identity
$D_{*i}B_{*i}=0$, $X=\alpha Y$ is a solution to the Gauss law
(\ref{jcteq}) for any $\alpha$. Since the $P$ term in $X^0$
(\ref{jctsol0}) generates this type of solution to the Gauss law,
we have only to consider the $Q$ term in (\ref{jctsol0}) as
$X^0$ on the RHS of (\ref{jcteq1}).
We shall put $\beta=1$ in (\ref{N}) for a while for the sake of
simplicity.
Then, the RHS of (\ref{jcteq1}) is evaluated by using the following
identities valid for $T_i$ of (\ref{Ti}):
\begin{eqnarray}
&&\{T_{ij},T_k\}=\delta_{ik}T_j+\delta_{jk}T_i
-\frac23\,\delta_{ij}T_k ,
\nn\\
&&\left[T_{ij},T_{kl}\right]
=i\left(\delta_{ik}\epsilon_{jlm}+\delta_{il}\epsilon_{jkm}
+\delta_{jk}\epsilon_{ilm} + \delta_{jl}\epsilon_{ikm}
\right)T_m .
\label{u3}
\end{eqnarray}
{}From the structure of the RHS of (\ref{jcteq1}), we see that $X^1$
consists of terms with one $T_i$ and even numbers of $\hx_i$.
Therefore the ansatz for $X^1$ is
\begin{eqnarray}
X^1=\bun{\xi r^3}\Bigl(
\theta_iT_i R(\xi)+\theta_i\hx_i\hx_jT_j S(\xi)\Bigr) .
\label{x1}
\end{eqnarray}
Note that we have factored out $1/\xi$ in (\ref{x1}) similarly to the
$Q$ term in (\ref{jctsol0}).
Putting the solution (\ref{mpsol0}) and the ansatz (\ref{x1}) into
the LHS of (\ref{jcteq1}), we obtain the following differential
equations for $R$ and $S$:
\begin{eqnarray}
&&\left(\calD^2 - 7\calD + 11 - H^2 + 2K -K^2\right)R
+ 2KS
\nn\\
&&\quad
= -\frac{23}{2}-\frac{21}{2}H +\frac52 H^2 +\frac12 H^3 -H^4
+\left(20 +32H +6 H^2 -6 H^3\right)K
\nn\\
&&\qquad
+\left(-\frac{19}{2}+ 9 H + 17 H^2 +\frac{23}{2}H^3 +3 H^4\right)K^2
+\left(-4 - 58 H- 52 H^2- 8 H^3\right)K^3
\nn\\
&&\qquad
+\frac12\left(35 + 75 H + 41 H^2\right)K^4
+\left(-16 -10 H\right)K^5 +\frac72 K^6
\nn\\
&&\qquad
+24z\left\{
(-1+3H^2+2H^3)K + 2(1+H-2H^2-2H^3)K^3 -(1+2H)K^5\right\} ,
\label{diffeqR}
\\[5pt]
&&\left(\calD^2 - 7\calD+10-2K-2K^2\right)S
+\left(-1 + H^2 + 2K - K^2\right)R
\nn\\
&&\quad
=\frac12\left(31 + 29H-5H^2 -H^3+2H^4\right)
+\left(-26 -34H - 2 H^2 + 6H^3\right)K
\nn\\
&&\qquad
+\left(\frac{19}{2}-23 H -23H^2 -\frac{15}{2}H^3-3 H^4\right)K^2
+\left(8 + 68H + 56H^2 + 8H^3\right)K^3
\nn\\
&&\qquad
-\frac12\left(43 + 71H + 41 H^2\right)K^4
+\left(18 + 10H\right)K^5 -\frac72 K^6
\nn\\
&&\qquad
+24z\Biggl\{
\left(1-3H^2-2H^3\right)K+\frac13\left(1-2H-3H^2+4H^3+4H^4\right)K^2
\nn\\
&&\qquad\qquad
+\left(-2-2H+4H^2+4H^3\right)K^3+\frac23\left(-1+H+2H^2\right)K^4
\nn\\
&&\qquad\qquad
+\left(1+2H\right)K^5+\frac13K^6
\Biggr\} .
\label{diffeqS}
\end{eqnarray}
Eqs.\ (\ref{diffeqR}) and (\ref{diffeqS}) are the coefficient of
$\theta_iT_i$ and $\theta_i\hx_i\hx_jT_j$ in (\ref{jcteq1}),
respectively.
The solution to these differential equations is again given as
polynomials of $H$ and $K$ as in the previous monopole
case (\ref{B})--(\ref{W}):\footnote{
There seems to be no physically meaningful homogeneous solution to
(\ref{jcteq1}) as far as we have examined using the polynomial
assumption (\ref{polynom}).
}
\begin{eqnarray}
&&R= -\frac32 -\bun{2}H + H^2
+\left(2 - H - 3H^2\right)K
+ \left(1+\frac52 H + H^2\right)K^2
\nn\\
&&\qquad
-(2+H)K^3+\bun{2}K^4
-8z\left(H^2 + H^3\right)K ,
\\[5pt]
&&S= \frac32 -\frac12 H -2 H^2
-\left(2 - H - 3H^2\right)K
-\left(1 +\frac32 H + H^2\right)K^2
+(2+H)K^3 -\frac12 K^4
\nn\\
&&\qquad
-8z\left\{
\frac12\left(1+H\right)^2 -\left(H^2 + H^3\right)K
-\left(1+H+H^2\right)K^2 +\frac12 K^4
\right\} .
\end{eqnarray}

Having obtained the classical solution for $X$, we shall evaluate its
eigenvalues. First, the zero-th order eigenvalues of $X$ are obtained
from (\ref{jctsol0}) with $(\alpha,\beta)=(0,1)$ as
\begin{eqnarray}
\lambda_X^0=\frac{2Q}{3\xi r}\pmatrix{1\cr -2\cr 1}
\sim\left(-\frac43C+\frac{2}{r}\right)\pmatrix{1\cr -2\cr 1} .
\label{lX0}
\end{eqnarray}
The $O(\theta)$ eigenvalue can be evaluated as in the previous
case using (\ref{eigenvalue}):
\begin{eqnarray}
\lambda_X^1=\frac{\theta_i\hx_i}{\xi r^3}
\Bigl(-Q+R+S\Bigr)
\pmatrix{1\cr 0\cr-1} .
\label{lX1}
\end{eqnarray}
Summing (\ref{lX0}) and (\ref{lX1}), the total eigenvalues
$\lambda_X=\left(\lambda_X^{(+)},\lambda_X^{(0)},\lambda_X^{(-)}
\right)^T$ of $X$ with $(\alpha,\beta)=(0,1)$
are given by
\begin{eqnarray}
&&\lambda_X^{(\pm)}\sim-\frac43C+\frac2r
\pm\frac{\theta\cdot\hx}{r^3}\Bigl((1-4z)\xi-2\Bigr)
\nn\\
&&\phantom{\lambda_X^{(\pm)}}
=-\frac43C
+2\left|x_i\pm\theta_i
\left(\frac12\lambda_X^{(\pm)}
+\left(\frac16+2z\right)C\right)\right|^{-1} ,
\label{lX(pm)}\\
&&\lambda_X^{(0)}\sim\frac83C-\frac{4}{r} .
\label{lX(0)}
\end{eqnarray}
String junction interpretation of these eigenvalues will be given in
the next section.

\section{Non-commutative string junction}

Now we would like to draw the string junction picture from the
asymptotic behavior of the eigenvalues of $X$ and $Y$ via
Callan-Maldacena interpretation.
First, the string junction picture projected on the $(X,Y)$ plane is
the same as in the $\theta=0$ case (Fig.\ \ref{XY}) since the
$O(\theta)$ corrections to the eigenvalues $\lambda_X$ and $\lambda_Y$
do not change their leading asymptotic behavior.

\begin{figure}[htb]
\begin{center}
\leavevmode
\epsfxsize=80mm
\put(-17,25){D3($\lambda_Y^{(+)}$)}
\put(-17,71){D3($\lambda_Y^{(0)}$)}
\put(-17,140){D3($\lambda_Y^{(-)}$)}
\put(205,25){$-C$}
\put(205,140){$C$}
\put(100,170){{\Large $Y$}}
\put(220,92){{\Large $x_i$}}
\put(160,15){{$C\theta_i$}}
\put(40,150){{$-C\theta_i$}}
\epsfbox{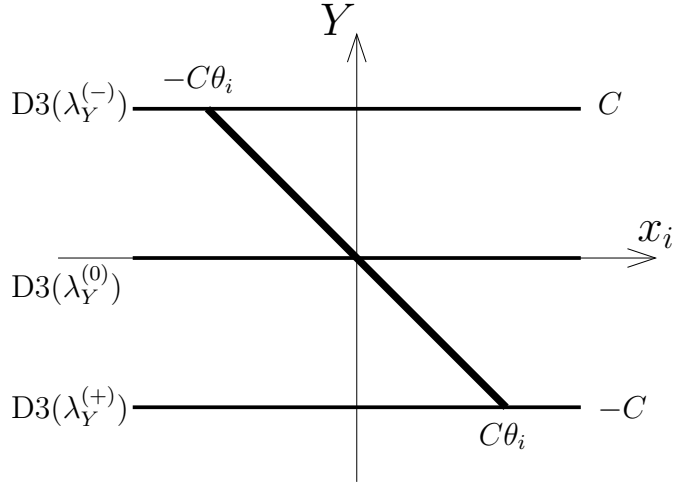}
\caption{String junction picture in the $(Y,x_i)$ space obtained
  from the asymptotic behavior of $\lambda_Y$. This is the special
  case of Fig.\ \ref{Y} with $z=-1/4$. The three $(p,q)$-strings are
  seen as one connected straight line.
}
\label{Yjct}
\end{center}
\end{figure}

Second, the string picture obtained in Sec.\ 3 from the eigenvalues
$\lambda_Y$ of (\ref{lY(pm)}) and (\ref{lY(0)}) gives the string
junction projected on the $(Y,x_i)$ space.
Since the string junctions should be connected, we have
to take the special value $z=-1/4$ (see Fig.\ \ref{Yjct}).
The three $(p,q)$-strings constituting the junction look as one
straight line tilted against the D3-branes by angle $\theta_i$.

Before examining the eigenvalues of $X$ for a general
$(\alpha,\beta)$, let us consider the brane interpretation of the
eigenvalues (\ref{lX(pm)}) and (\ref{lX(0)}) corresponding to
$(\alpha,\beta)=(0,1)$. In this case, we obtain a three-string
junction picture where each of the three strings is attached to the
respective D3-brane at the points $x_i=\pm(1-4z)C\theta_i/2$ and
$x_i=0$ for the branes corresponding to $\lambda_X^{(\pm)}$ and
$\lambda_X^{(0)}$, respectively.
Surprisingly, these endpoint $x_i$ coordinates of the three strings in
the $(X,x_i)$ space coincide with the corresponding ones in the
$(Y,x_i)$ space only when $z=-1/4$, and they are given by
$x_i=\pm C\theta_i$ and $x_i=0$.
In the following we shall restrict our arguments to the case $z=-1/4$.

\begin{figure}[htb]
\begin{center}
\leavevmode
\epsfxsize=80mm
\put(-17,25){D3($\lambda_X^{(0)}$)}
\put(-17,140){D3($\lambda_X^{(+)}$)}
\put(-17,165){D3($\lambda_X^{(-)}$)}
\put(95,195){{\Large $X$}}
\put(220,120){{\Large $x_i$}}
\put(35,175){{ $-C\theta_i$}}
\put(160,147){{$C\theta_i$}}
\put(120,86){{$2\beta C/3$}}
\put(202,25){{$8\beta C/3$}}
\put(202,140){{$(-\alpha-4\beta/3)C$}}
\put(202,170){{$(\alpha-4\beta/3)C$}}
\epsfbox{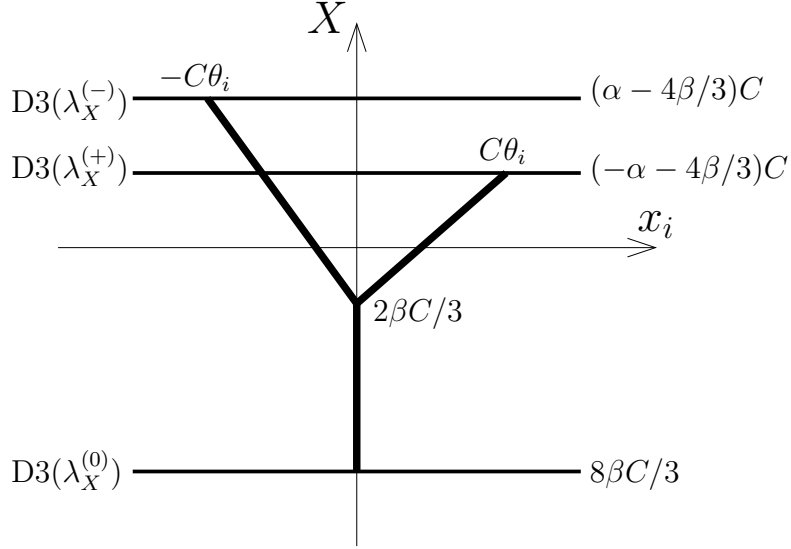}
\caption{String junction picture in the $(X,x_i)$ space obtained
  from the asymptotic behavior of $\lambda_X$, (\ref{lX(pm)ab}) and
(\ref{lX(0)ab}), with $z=-1/4$. This figure represents the case
with $\alpha>0$ and $\beta<0$.
}
\label{X}
\end{center}
\end{figure}

For a general $(\alpha,\beta)$, the eigenvalues of $X$ are given as
$\alpha\lambda_Y + \beta\times\left[
\lambda_X\mbox{ of (\ref{lX(pm)}) and (\ref{lX(0)})}\right]$.
Explicitly, they are
\begin{eqnarray}
&&\lambda_X^{(\pm)}
\sim\left(\mp\alpha-\frac43\beta\right)C
+\left(\pm\alpha+2\beta\right)\left|x_i
\pm\frac{\theta_i}{\pm\alpha+2\beta}
\left(\lambda_X^{(\pm)}-\frac23\beta C\right)\right|^{-1} ,
\label{lX(pm)ab}
\\
&&\lambda_X^{(0)}\sim
\frac83\beta C- \frac{4\beta}{r} .
\label{lX(0)ab}
\end{eqnarray}
The string picture of these eigenvalues is given in Fig.~\ref{X}.

Let us summarize various quantities of the three-string junction
picture obtained here. First, the endpoint coordinates of the three
strings in the $(X,Y,x_i)$ space are
\begin{eqnarray}
C\left(\mp\alpha-\frac43\beta,\mp 1,\pm\theta_i\right),\quad
C\left(\frac83\beta,0,0\right) ,
\end{eqnarray}
for the strings $(\pm)$ and $(0)$, respectively, and the three strings
meet at the point
\begin{equation}
C\left(\frac{2\beta}{3},0,0\right) .
\end{equation}
Defining the $(p,q)$-charges of the strings by the leading
asymptotic behavior of the eigenvalues of the electric and magnetic
fields as
\begin{eqnarray}
\left(E_i,B_i\right)\sim \frac{\hx_i}{2r^2}\,(p,q) ,
\end{eqnarray}
we have
\begin{eqnarray}
\left(p,q\right)^{(\pm)}
=\left(\mp 2\alpha-4\beta,\mp 2\right) ,
\quad
\left(p,q\right)^{(0)}=\left(8\beta,0\right) .
\end{eqnarray}
Then, the tension vectors are given by
\begin{eqnarray}
\vec T=(p,q,-q\,\theta_i) ,
\end{eqnarray}
for each of the three strings, and they are balanced,
$\sum \vec{T}=0$.

\begin{figure}[htb]
\begin{center}
\leavevmode
\epsfxsize=80mm
\put(100,193){{\Large $Y$}}
\put(0,57){{\Large $X$}}
\put(220,105){{\Large $x_i$}}
\put(75,25){{\large $\theta_i$}}
\put(-6,118){{\large D3}}
\put(200,125){{\large D3}}
\put(150,37){{\large D3}}
\epsfbox{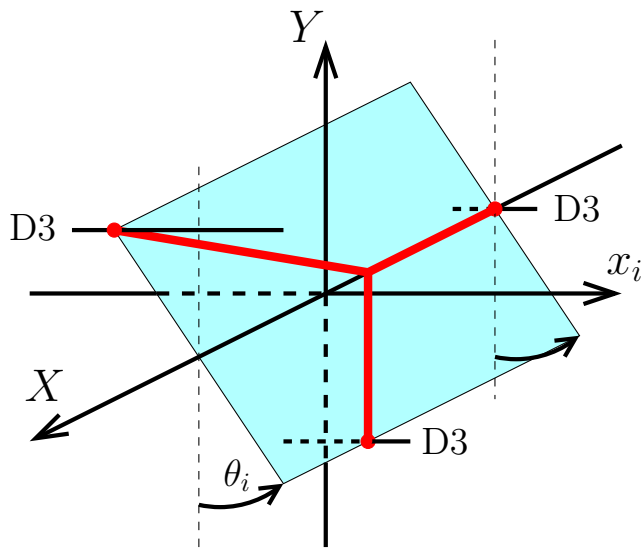}
\caption{The non-commutative string junction (on the shaded plane) is
  obtained from the ordinary one with $\theta=0$ by a rotation
  around the $X$-axis with rotation angle $\theta_i$.
}
\label{3D}
\end{center}
\end{figure}

As seen from the above analysis,
the present string junction with nonvanishing $\theta$ is obtained
from that with $\theta=0$ (which is on the $x_i=0$ plane) by a
rotation around the $X$-axis with angle $\theta_i$
(see Fig.\ \ref{3D}).
These results are consistent with the expectation obtained from the
force balance among the string tension and the magnetic force, which
is felt by the charge $q$ at the endpoint of each string in the
uniform magnetic field $\theta_i$ \cite{HH}.

\section{Summary and discussion}

In this paper we have constructed a $1/4$ BPS soliton solution in
${\cal N}=4$ non-commutative super Yang-Mills theory. From the
asymptotic behavior of the scalar eigenvalues, we have successfully
reproduced the expected string junction picture.
We would like to emphasize that such consistent eigenvalues can never
be obtained without the Poisson bracket term in the eigenvalue formula
(\ref{eigenvalue}). Thus
our results give further support of the non-commutative eigenvalue
equation (\ref{Mv=lv}) proposed in ref.\ \cite{HHM}.

There are a number of questions to be clarified. First, we have chosen
the special value $z=-1/4$ as the parameter specifying the separation
of the two monopole centers. This led to a consistent string junction
picture. However, we have constructed a solution for any value of
$z$, and a question is what the string theory interpretation of our
solution for $z\ne -1/4$ is.
A similar problem exists already in the $1/4$ BPS solution in the
ordinary super Yang-Mills theory with the moduli of the separation of
the monopole centers \cite{LY}.

Another question is the simultaneous diagonalizability of the two
scalars $X$ and $Y$. In the case $\theta=0$, our spherically symmetric
solution satisfies $[X^0,Y^0]=0$ and hence we can consider the
eigenvalues  of $X^0$ and $Y^0$ simultaneously.
However, in the non-commutative case, the eigenvectors $\vv$ of the
eigenvalue equation (\ref{Mv=lv}) are generally different for $X$ and
$Y$. We have to justify the present analysis where we considered the
eigenvalues of both $X$ and $Y$. It is expected that our analysis at
the asymptotic region $r\to\infty$ is valid since $\theta$ is always
multiplied by negative powers of $r$.

\section*{Acknowledgments}
We would like to thank T.~Asakawa, S.~Goto, K.~Hashimoto, S.~Iso
and Y.~Yoshida for valuable discussions and comments.
This work is supported in part by Grant-in-Aid for Scientific Research
from Ministry of Education, Science, Sports and Culture of Japan
(\#09640346, \#04633).
The work of S.\ M.\ is supported in part by the Japan Society for the
Promotion of Science under the Predoctoral Research Program.

\end{document}